# Downscaling of non van der Waals Semimetallic W$_5$N$_6$ with Resistivity Preservation

*Hongze Gao[1], Da Zhou[2], Lu Ping[3], Zifan Wang[1], Nguyen Tuan Hung[4], Jun Cao[1], Michael Geiwitz[5], Gabriel Natale[5], Yuxuan Cosmi Lin[6], Kenneth Stephen Burch[5], Riichiro Saito[7,8], Mauricio Terrones[2,9-11], and Xi Ling* [1,3]*

[1]Department of Chemistry, Boston University

590 Commonwealth Ave., Boston, MA, 02215, U.S.

[2]Department of Physics, The Pennsylvania State University

104 Davey Laboratory, University Park, PA, 16802, U.S.

[3]Division of Materials Science and Engineering, Boston University

15 St Mary's St., Boston, MA, 02215, U.S.

[4]Frontier Research Institute for Interdisciplinary Sciences, Tohoku University

Sendai, 980-8578, Japan

[5]Department of Physics, Boston College

140 Commonwealth Ave., Chestnut Hill, MA, 02467, U.S.

[6]Department of Materials Science and Engineering, Texas A&M University

575 Ross St., College Station, TX, 77843, U.S.

[7]Department of Physics, Tohoku University

Sendai 980-8578, Japan

[8]Department of Physics, National Taiwan Normal University

Taipei 11677, Taiwan

[9]Department of Chemistry, The Pennsylvania State University

77 Pollock Rd, State College, PA, 16801, U.S.

[10]Department of Materials Science and Engineering, The Pennsylvania State University

116 Deike Building, University Park, PA, 16802, U.S.

[11]Center for 2-Dimensional and Layered Materials, The Pennsylvania State University Pollock Rd, University Park, PA, 16802, U.S.

Email: xiling@bu.edu

**Abstract**

The bulk phase of transition metal nitrides (TMNs) has long been a subject of extensive investigation due to their utility as coating materials, electrocatalysts, and diffusion barriers, attributed to their high conductivity and refractory properties. Downscaling TMNs into two-dimensional (2D) forms would provide valuable members to the existing 2D materials repertoire, with potential enhancements across various applications. Moreover, calculations have anticipated the emergence of uncommon physical phenomena in TMNs at the 2D limit. In this study, we use the atomic substitution approach to synthesize 2D $W_5N_6$ with tunable thicknesses from tens of nanometers down to 2.9 nm. The obtained flakes exhibit high crystallinity and smooth surfaces. Electrical measurements on 15 samples show an average electrical conductivity of 161.1 S/cm, which persists while thickness decreases from 45.6 nm to 2.9 nm. The observed weak gate tuning effect suggests the semimetallic nature of the synthesized 2D $W_5N_6$. Further investigation into the conversion mechanism elucidates the crucial role of chalcogen vacancies in the precursor for initiating the reaction and strain in propagating the conversion. Our work introduces a desired semimetallic crystal to the 2D material library with mechanistic insights for future design of the synthesis.

**Introduction**

Transition metal nitrides (TMNs) have been studied for decades owing to their excellent physical and chemical properties such as high electrical and thermal conductivity,[1,2] high hardness,[2–4] and outstanding refractory nature.[5] Taking advantage of these excellent properties, TMNs have been used in various applications including electrocatalysts,[6] energy storage,[7,8] and diffusion barriers.[9,10] Despite the demands of these properties in two-dimensional (2D) materials family for robust and high performance electronics,[11,12] downscaling TMNs into 2D form remains a challenging task. Majority of TMNs are non van der Waals (vdW) crystals with strong interaction along all directions across the lattice,[13,14] imposing technical difficulties in creating 2D layers from a bulk crystal. Conventional bottom-up techniques including chemical vapor deposition (CVD), magnetron sputtering, and molecular beam epitaxy (MBE) have been used to prepare thin films of TMNs,[1,15] but the synthesis of few-nm thick films has been limited by rough surface due to the island-like growth mechanism.[16,17] Intriguingly, recent theoretical studies predicted new properties in the 2D form of tungsten nitrides ($WN_x$, where the subscript x refers to different stoichiometries) compared to their bulk counterparts. For example, Chen *et al.* and Campi *et al.* predicted the emergence of nontrivial topological states in 2D $WN_x$.[18,19] Namely, 2D $W_3N_4$ and $W_2N_3$ are predicted to exhibit topological superconductivity with critical temperature as high as 11 K and 21 K, respectively.[18,19] Chen *et al.* also predicted the presence of nodal lines in 2D $W_3N_4$ that were robust against spin-orbital coupling, which was uncommon in bulk $WN_x$ or any other materials.[18] Chin *et al.* and Zhao *et al.* further showed 2D $W_5N_6$ to be a semimetal, which has low density of states (DOS) around Fermi level ($E_F$).[14,20] This last property is particular interesting for 2D semiconductor contacts and for future interconnects. Indeed, Shen *et al*. recently reported that using semimetallic bismuth as contact electrodes material to 2D semiconductors efficiently reduces metal-induced gap states (MIGS) at the metal-semiconductor interfaces,[21] and there is mounting evidence that topological semimetals can be useful for enhanced electronic transport at the nanoscale.[22,23] Therefore, experimentally realizing 2D $WN_x$ will open up tremendous opportunities to explore their exotic quantum properties and applications in electronics.

To this end, a few methods have been reported to obtain 2D $WN_x$. For instance, WN, $W_2N$, and $W_2N_3$ nanosheets have been achieved through a template-assisted ammonization process on various compounds containing tungsten;[24–26] ultra-thin films of $W_5N_6$ have been obtained through a vapor-liquid-solid growth process;[20,27] 2D WN flakes have been synthesized through a salt-assisted CVD approach.[28] A few applications including supercapacitors,[25] surface-enhanced Raman spectroscopy (SERS),[26] and electrochemical catalysis have been demonstrated with excellent performances using the synthesized samples.[24,29,30] Moreover, Chin *et al*. reported improved performance of $MoS_2$ field effect transistors (FETs) with 2D $W_5N_6/MoS_2$ contact compared to the conventional $Au/MoS_2$ contact.[20] However, the 2D $WN_x$ prepared in these works usually possess a rough and non-uniform surface and small grain sizes. The limited sample lateral sizes (< 5 µm) foster technical difficulties in device fabrication and hinder the measurements on individual crystalline flakes. The rough surface also prohibits the effective integration between the samples and other 2D materials due to weak adhesion. Furthermore, rough surface causes severe charge scattering especially at the 2D limit and raises the electrical resistivity of the material.[31] This also restricts the miniaturization of conventional interconnect materials for nanoelectronic applications.[32]

Very recently, an atomic substitution approach has been developed to obtain high quality 2D TMNs for the study of nanoelectronics. Cao *et al.* demonstrated the conversion of 2D $MoS_2$ into 2D $Mo_5N_6$, where the 2D nature, high crystallinity, and smooth surface of $MoS_2$ is preserved in $Mo_5N_6$.[33] While it is promising to realize other metal nitrides using this approach,[33–35] the investigation on the atomic substitution of tungsten dichalcogenides for $WN_x$ is very limited. In this work, we investigate the nitrogen substitution of $WX_2$ (X=S, Se and Te). 2D $W_5N_6$ with the desired quality for nanoelectronics is obtained using $WSe_2$ as a precursor in the atomic substitution process. The surfaces of the obtained 2D $W_5N_6$ are as smooth as that of the $WSe_2$ precursor, which plays a crucial role in the persistence of its relatively high electrical conductivity down to sample thicknesses of 2.9 nm. A weak gate tuning effect of conductance indicates the semimetallic nature of synthesized 2D $W_5N_6$. Moreover, by comparing the results from using different tungsten-based transition metal dichalcogenides (W-TMDs) as precursors, we show a positive correlation between the reaction threshold temperature and the formation energies of chalcogen vacancies. A mechanistic understanding of the conversion process, highlighting the importance of chalcogen vacancies in initiating the process and the strain in driving the reaction, is further proposed.

**Results and Discussion**

Figure 1a shows side and top views of the crystal structure of WSe$_2$ precursor and the converted W$_5$N$_6$. Unlike WSe$_2$, vdW gaps are missing in W$_5$N$_6$, and chemical bonds extend along all three dimensions instead. The W$_5$N$_6$ crystal is composed of W atoms in a hexagonal close-packed structure, with N atoms occupying the interstitial sites.[14] If all interstitial sites are occupied by N atoms, the WN phase will form. In the structure of W$_5$N$_6$, a W vacancy appears in one of every six W sites, resulting in W$_5$N$_6$ phase with a non-unity stoichiometry of W and N. Owing to the elimination of vdW gaps, the distance between two adjacent W layers (along the c-axis) decreased from 7.53 Å to 2.82 Å. Thus, the thickness of the flake is expected to decrease to 37.5% of its original value after the conversion process.

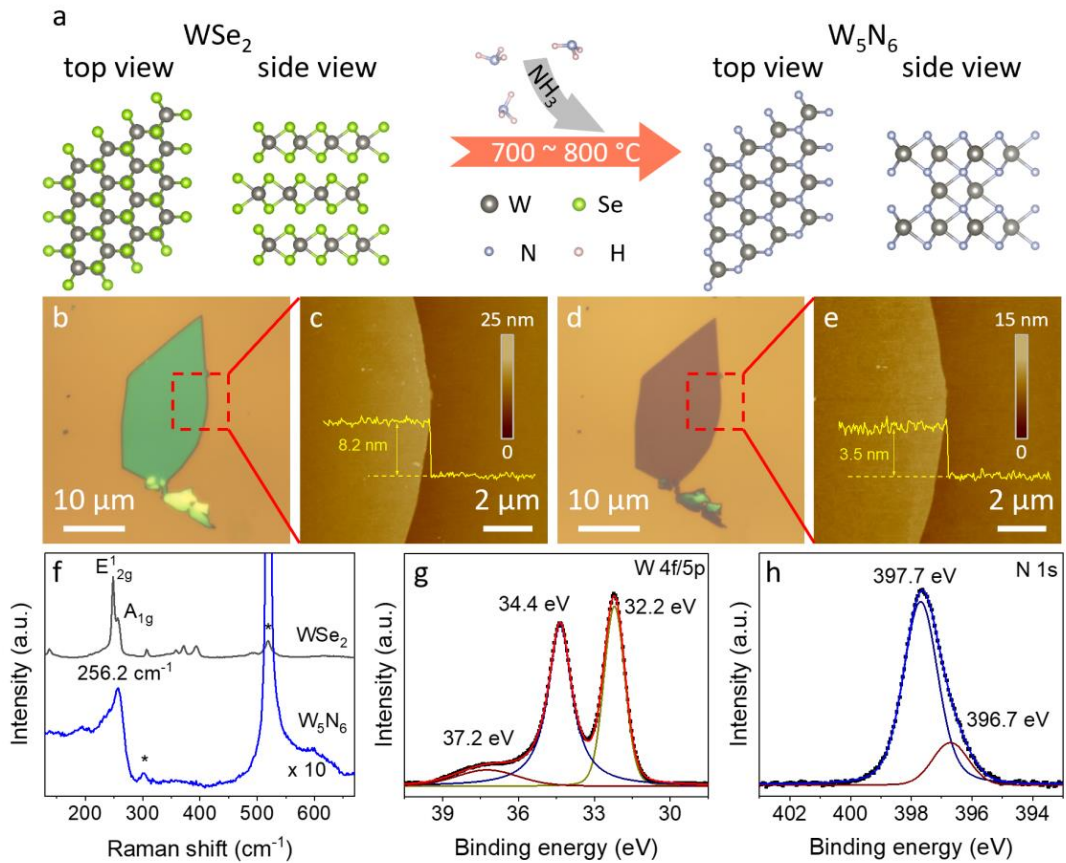

**Figure 1 Crystal structure, optical and AFM images, and spectroscopy characterizations of 2D W$_5$N$_6$.** (a) Illustration of the conversion from WSe$_2$ to W$_5$N$_6$ through a nitridation reaction. (b, d) Optical images of the WSe$_2$ and converted flake. (c, e) AFM image of the area labeled by red squares in (b) and (d). (f) Raman spectra of WSe$_2$ precursor and the converted flake. (g, h) W and N orbital of XPS spectrum measured on converted flakes.

As shown in Figures 1b and 1d, a substantial change in optical contrast is observed in the flake after the reaction due to the distinct optical properties of the $W_5N_6$ from $WSe_2$, consistent with our previous observation on the $MoS_2$-to-$MoN_x$ conversions.[33,35] Atomic force microscopy (AFM) images of the flakes show a smooth surface inherited from the $WSe_2$ precursors (Figure 1c and 1e). The arithmetic average roughness ($R_a$) of $W_5N_6$ (0.25 nm) is comparable with that of the $WSe_2$ precursors (0.26 nm). AFM measurements also show that the thickness of the flake decreased from 8.2 nm ($WSe_2$) to 3.5 nm ($W_5N_6$), corresponding to the vanishing of vdW gaps between each atomic layer. The experimental thickness decrement of 42.7% also matches well with the theoretical value of 37.5%.

To characterize the composition of the converted flakes, we perform Raman spectroscopy and X-ray photoelectron spectroscopy (XPS) characterizations. Figure 1f shows the typical Raman spectrum of converted flake measured on a 16.1 nm thick sample and the comparison with the Raman spectrum of the $WSe_2$ precursor. Raman signatures from the $SiO_2$/Si substrate are labelled with an asterisk. The characteristic peaks of $WSe_2$ at 248.6 and 255.9 cm$^{-1}$ vanished after the reaction,[36] whereas a new peak centred around 256 cm$^{-1}$ with weaker intensity arouse after the conversion. Note that a plateau appears on the lower wavenumber side of the emerging Raman peak in the converted samples, which indicates the presence of multiple Raman bands in this range as supported by our phonon dispersion calculation in Figure S1a. Multiple phonon bands between 150 cm$^{-1}$ and 200

cm$^{-1}$ are present according to the calculation. The arising Raman peak matches the Raman signature of W$_5$N$_6$ reported in other works.[20,33]

Apart from WSe$_2$, we also performed the nitridation reaction on WS$_2$ and WTe$_2$. To distinguish the conversion on different W-TMDs, we label the converted flakes as WN$_x$-S, WN$_x$-Se, and WN$_x$-Te for WS$_2$, WSe$_2$, and WTe$_2$ precursors, respectively. We observed three characteristic Raman peaks in the converted WN$_x$ flakes, whose intensities vary with the reaction conditions and precursors (Figure S2). Given the rich isomers of WN$_x$ and their close formation energies,[14,37] we attribute the variable intensities of Raman peaks to the phase transitions between the isomers of WN$_x$. More details are provided in the supplementary information. We also noticed that the threshold temperature of the reaction varies with precursors. We will discuss this in a later section. The conversion of WSe$_2$ to W$_5$N$_6$ is further confirmed with XPS and transmission electron microscopy (TEM) described below.

An XPS survey of the synthesized flakes confirms the complete substitution reaction as no signals from Se are observed, while signals from N appear (see Figure S3). High-resolution XPS shown in Figures 1g and 1h clearly demonstrates strong signals from W and N orbitals after conversion. The W spectrum can be fitted into three peaks, *i.e.*, W 4f$_{7/2}$ (32.2 eV), W 4f$_{5/2}$ (34.4 eV), and W 5p$_{3/2}$ (37.2 eV). The W 4f$_{7/2}$ peak position is slightly lower than the values reported on WS$_2$ (32.8 eV), WSe$_2$ (32.6 eV) and WO$_2$ (33.0 eV), indicating an oxidation state slightly lower than +4 in the synthesized flakes.[38–40] This matches with the average oxidation state of +3.6 for W in W$_5$N$_6$. Note that the N 1s core level signal fits well with two instead of just one peak, with a strong peak centered at 397.7 eV and a secondary peak at 396.7 eV. The strong peak is attributed to W-N bonding,[20,41] and the secondary peak at lower binding energy is attributed to nitrogen adsorbates on the surface of the flakes,[42] which is possibly introduced during the synthesis process.

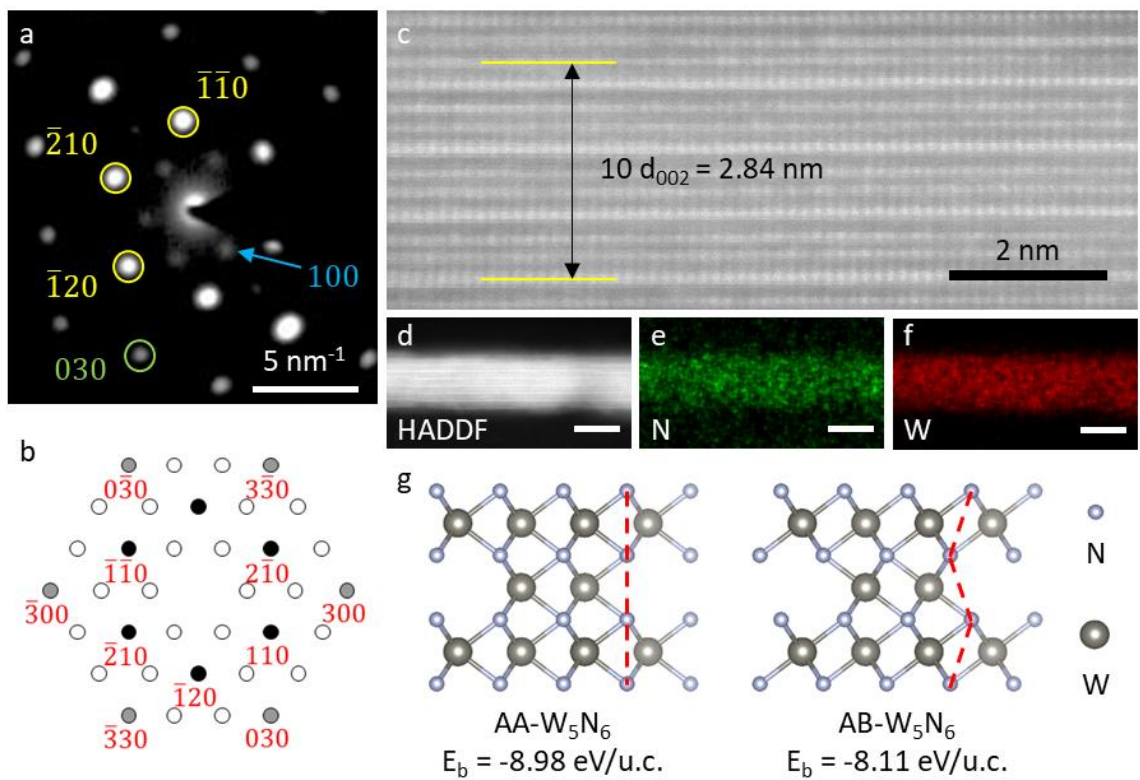

**Figure 2 TEM characterization and structure simulations.** (a) SAED of 2D $W_5N_6$ crystal taken along ⟨001⟩ direction. (b) Simulated SAED pattern of a 28 nm thick $W_5N_6$ with incident electron beam along [001] crystal direction. Black and grey spots represent predicted diffraction spots with high and low intensity, respectively. (c) Cross-sectional HAADF-STEM image. The distance is measured across ten W atomic layers along ⟨001⟩ direction. (d-f) Low magnification HAADF (d), N (e) and W (f) element mapping of $W_5N_6$ in cross-sectional STEM image. The scale bars are 5 nm. (g) Structure of two phases (*i.e.*, AA and AB) of $W_5N_6$ with different stacking orders of N atoms. The calculated binding energy of each phase is shown.

To characterize the crystal structure of synthesized flakes, we perform TEM measurements. The low-magnification TEM of a $W_5N_6$ flake is shown in Figure S4. The selected area electron diffraction (SAED) pattern (Figure 2a) of planar TEM image shows sharp diffraction points with hexagon patterns, suggesting the high crystallinity of converted flakes. As label in Figure 2a, a sharp set of diffraction patterns from different crystal planes is observed, matching with the simulation results with incident electrons along the [001] crystal orientation of $W_5N_6$ (Figure 2b). A wavevector value of 4.34 nm$^{-1}$ is measured from the smallest set of bright diffraction spots (labelled as ($\bar{1}\bar{1}0$)), corresponding to the interplanar distance of 2.30 Å. This matches with the interplanar distance between (110) planes ($d_{110}$) of $W_5N_6$ (*i.e.*, 2.45 Å).[14] It is worth noting that another set of weak spots are observed with smaller wavevectors, attributed to the {100} planes from simulation results. As $W_5N_6$ structure can be viewed as the WN structure with periodic W vacancies, we also simulate the SAED pattern for defect-free WN crystals (Figure S5), where the {100} diffraction spots are missing. Hence the emergence of {100} diffraction spots serve as evidence for W vacancies. Secondary diffraction patterns from the {300} group are also observed with lower intensity, matching with the simulation results.

In addition to the planar TEM, we also perform cross-sectional scanning TEM (STEM) characterization on the obtained 2D $W_5N_6$ flakes. Individual layers of W atoms can be clearly recognized in Figure 2c via high-angle annular dark field (HAADF)-STEM imaging. The average distance between two adjacent W layers along the c-axis is extracted to be 2.84 Å, which is in good agreement with the theoretical value (*i.e.*, 2.82 Å).[14] This value is 42.4% of the corresponding value in $WSe_2$ (6.7 Å), resulting from elimination of the vdW gap.[43] We also show the elemental mapping of W and N in a cross-sectional image (Figure 2d-2f). N and W are uniformly distributed across the atomic layers. The results further confirm the synthesized flakes as $W_5N_6$ with high crystallinity.

Although STEM images provide clear measurements of the spatial distribution of W atoms, N atoms are not resolved due to the weak signal intensity from their low atomic number. To understand the occupancy of interstitial N atoms, we calculate the binding energies of two different phases of $W_5N_6$, where they differ from each other in the coordinate environment of the $[WN_6]$ complexes. We label these two phases as AA-$W_5N_6$ (trigonal prismatic $[WN_6]$ complex) and AB-$W_5N_6$ (octahedral $[WN_6]$ complex), and show the side views of the calculated structures in Figure 2g. In AA-$W_5N_6$, the N atoms in the top layer of the unit cell are located right above the N atoms in the bottom layer, while a misalignment of the top and bottom N atomic layers occurs in AB-$W_5N_6$. The red dashed lines in Figure 2g distinguish the AA and AB stacking order. Our phonon dispersion calculations (Figure S1) of AA-$W_5N_6$ and AB-$W_5N_6$ show that both are dynamically stable structure, as there are no phonon bands with negative frequency. We further calculate the binding energy ($E_b$, defined as the energy change during the formation of crystal from individual atoms) of both phases with density functional theory (DFT), where the $E_b$ of AA-$W_5N_6$ (-8.98 eV/unit cell) is found lower than that of AB-$W_5N_6$ (-8.11 eV/unit cell). Hence, the AA stacking order of interstitial N atoms is thermodynamically more stable in $W_5N_6$.

To investigate the electrical properties of as-synthesized 2D $W_5N_6$, we fabricate 5/45 nm Cr/Au electrodes on 15 $W_5N_6$ flakes with different thicknesses using a cleanroom in a glovebox to minimize contaminants.[44] As shown in Figure 3a, four-point measurements are performed to eliminate the influence of contact resistances. Figure 3b shows the back-gate measurements configuration. Typical I-V characteristic curves of samples with different thicknesses are presented in Figure 3c, where a linear trend was observed, indicating the formation of Ohmic contact between $W_5N_6$ and Cr/Au electrodes.[35] We further extract electrical conductivity of each sample after considering the dimension factors (*i.e.*, channel length, width, and height). As shown in the inset of Figure 3c, the electrical conductivity of $W_5N_6$ is 161.1 ± 42.1 S/cm.

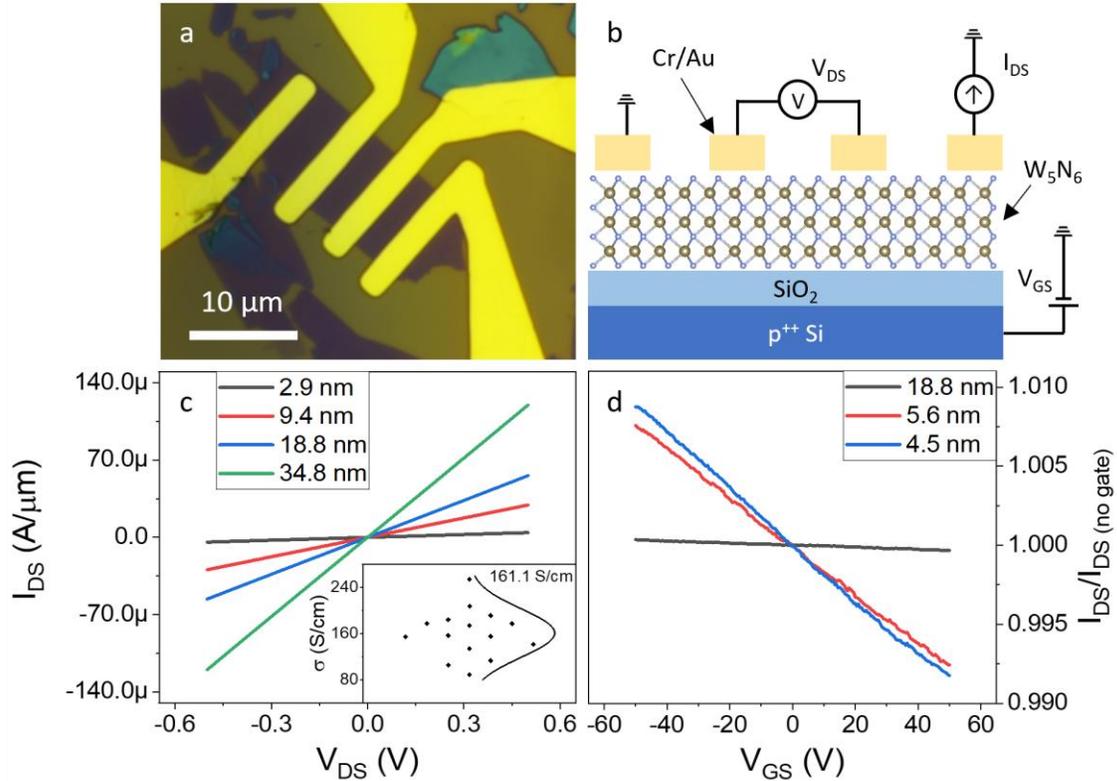

**Figure 3** Electrical measurements. (a, b) Optical image (a) and schematic drawing (b) of $W_5N_6$ with Cr/Au electrodes. (c) I-V characteristic curves of $W_5N_6$ of various thicknesses with no gate voltages. Inset presents the electrical conductivity extracted from each sample. (d) Transfer curves of $W_5N_6$ with various thicknesses. $V_{DS}$ = 0.5 V. $I_{DS}$ is normalized to the corresponding values with $V_{GS}$ = 0 V.

Moreover, we performed transfer characteristics measurements with application of a gate voltage ($V_g$) to verify the semimetallic properties of synthesized 2D $W_5N_6$. A weak tuning effect of conductance is observed as shown in Figure 3d. Three typical transfer characteristic curves measured on different thicknesses are presented. Drain-source current ($I_{DS}$) decreases from 12.08 µA at $V_{GS}$ = -50 V to 11.88 µA at $V_{GS}$ = +50 V on a 4.5-nm-thick sample, corresponding to the ratios of 1.009 and 0.992 in comparison to $I_{DS}$ at $V_{GS}$ = 0 V. The decrement in current at higher electron doping (*i.e.*, positive $V_{GS}$) indicates that the charge transport in 2D $W_5N_6$ is dominated by positive charge carriers (*i.e.*, holes).[45] The weak gate tunability is attributed to the semimetallic nature (low DOS at $E_F$) of $W_5N_6$, which has been predicted and verified in previous reports.[14,20] We also observe a decreased gate tuning on thicker samples, which is attributed to the screening effect from the bottom layers to top layers of $W_5N_6$, similar to that reported on graphene and $MoS_2$.[46,47] Taking advantage of the semimetallic properties of $W_5N_6$, we fabricate $MoS_2$ transistors with $W_5N_6$ contacts and find a much lower resistance compared to Au contacts (Figure S6)

It is worth noting that the conductivity values of synthesized $W_5N_6$ samples exhibit small fluctuation during the downscaling of thicknesses from 45.6 nm to 2.9 nm (Figure S7). Such scaling trend is distinct from metal thin films prepared by vacuum deposition. For example, the electrical conductivity of Cu dropped by five orders of magnitude from $1.4 \times 10^5$ S/cm to 0.84 S/cm when thickness decreased from 12.3 nm to 4.2 nm (Figure S7).[31] The degradation of electrical conductivity in metal thin films is usually due to: (1) prominent surface scattering with low film thickness,[32] (2) smaller grain size and consequentially more frequent grain boundary scattering at reduced thicknesses,[32,48] and (3) introduction of voids during deposition, which significantly decreases the scattering time and reduces the mean free path of charge carriers.[49] Therefore, the electrical resistivity of ultra-thin films with thicknesses lower than the critical thickness (typically below 10 nm) is usually much higher than their bulk values.[31,49] To mitigate the scaling effect on resistivity, one approach is to prepare sample with smooth surface to alleviate the surface scattering.

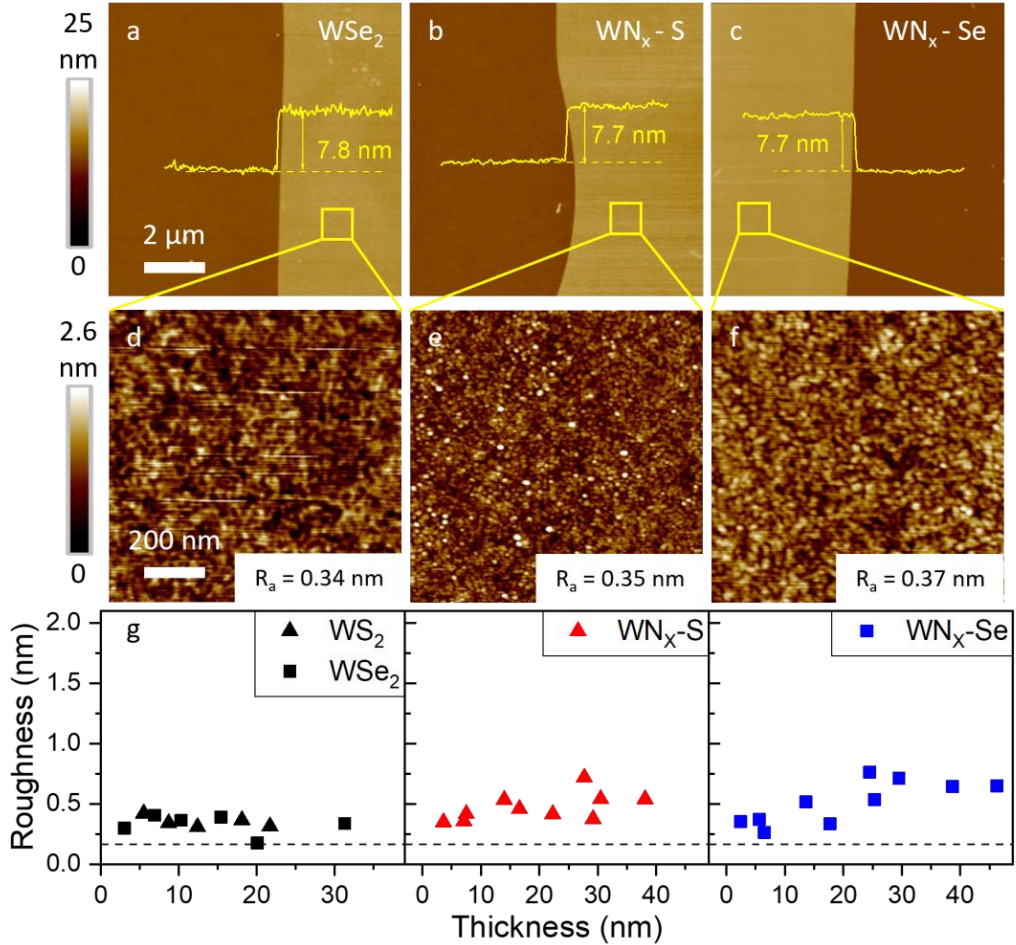

**Figure 4 Surface roughness measurements.** (a-c) Large area (10 μm × 10 μm) AFM scans of TMDs precursor (a), $WN_x$-S (b) and $WN_x$-Se (c). Image in (a-c) share the same scale bar. (d-f) Zoom-in AFM scans (1 μm × 1 μm) of the areas in (a-c) labeled by the yellow square. Image in (d-f) share the same scale bar. (g) Summary of roughness of precursors and $WN_x$.

Hereby, we perform AFM measurements on $WN_x$ of various thicknesses to investigate their surface roughness. We also measured the $SiO_2$/Si surface as a reference (Figure S). Figure 4a-4c show the AFM images of a large area (*i.e.*, 10 μm × 10 μm), where smooth surfaces are observed on $WSe_2$ precursor, $WN_x$-S, and $WN_x$-Se. Images of small areas (1 μm × 1 μm) are shown in Figure 4d-4f for the roughness measurements. Figure 4g summarizes the $R_a$ value of each sample. It is worth noting that for thin (< 20 nm) $WN_x$, the surface is smooth ($R_a$ = 0.39 ± 0.08 nm) and comparable to the TMDs precursors ($R_a$ = 0.34 ± 0.07 nm). For $WN_x$ thicker than 20 nm, the surface roughness ($R_a$ = 0.61 ± 0.12 nm) is much higher than thinner flakes. This can be attributed to excessive strain induced by lattice reconstruction during the reaction.[50] Note that all $R_a$ values are well below that of Cu film deposited at ultra-high vacuum ($R_a$ = 1.82 nm at thickness of 60 nm).[51] Cu films usually exhibit island-like growth in the sub-10-nm regime during deposition,[16] which produce rough surface and significant thickness fluctuations that deteriorates their electrical properties. Benefiting from the smooth surface, charge scattering is significantly reduced and the resistivity of the synthesized $W_5N_6$ flakes does not exhibit drastic increase at the ultra-thin regime (Figure S7). We also measured the roughness of $WN_x$-Te (Figure S9) and extracted a much higher $R_a$ value (1.54 ± 0.66 nm) than $WN_x$-S and $WN_x$-Se. This is probably due to the large lattice mismatch between $WTe_2$ and $WN_x$ and hence a large distortion during the reaction.[52]

Atomic substitution has been demonstrated as an effective approach to obtain non vdW 2D crystals, including GaN,[34] $Mo_5N_6$,[33,35] $Mo_2C$,[53] $InF_3$,[54] MoP,[50] CdS,[55] and several metal oxides.[56–58] The method also allows the fabrication of vertical and lateral heterostructures between vdW and non vdW 2D materials with clean interfaces.[57,59,60] Nevertheless, there is little study focusing on understanding how the substitution works at atomic level. As reported in previous studies, the metal skeleton remains during the substitution reaction, while the chalcogen elements are replaced with other elements (N, C, F, etc.).[33,50,53,54] Li *et al*. revealed from *in-situ* TEM characterizations that the substitution reaction can initiate at both edge and surface of an $MoS_2$ flake,[59] which is attributed to dangling bonds and S vacancies on the surface. Hereby, through the comparison investigation of the reactivity of various W-TMDs (*i.e.*, $WS_2$, $WSe_2$ and $WTe_2$) in the conversion process, we gain an in-depth understanding on the mechanism of this substitution reaction.

Optical images of precursors and converted flakes are shown in Figure S10. We observe that the threshold temperature to initiate the reaction ($T_{th}$) decreases from $WS_2$ (740 °C) to $WSe_2$ (640 °C) and gets lower with $WTe_2$ (500 °C). This suggests increasing reactivity with $NH_3$ while chalcogen element goes downward along the periodic table in W-TMDs. Furthermore, we perform the reaction on all three TMDs at the same temperature (Figure 5a-f). At 700 °C, $WTe_2$ is fully converted in 30 minutes, while $WSe_2$ is partially converted, and no signs of conversion are observed for $WS_2$ within the same time scale as shown in Figure S11. This suggests that the reaction rate increases as chalcogen element goes downward.

To interpret the impact of chalcogen elements on the reactivity, we propose a defect-initiated and strain-driven mechanism in a microscopic picture, as demonstrated in Figure 5a: (i.-ii.) Creation of a chalcogen vacancy ($V_x$). Chalcogen atom gains excess energy at high temperature to break the W-chalcogen bonds and leave a $V_x$ at its lattice site. (ii.-iii.) Formation of W-N bonds. The $V_x$ serve as an active site for $NH_3$ to react with the W-TMDs precursors. A nitrogen atom occupies the $V_x$ site and bond with adjacent W atoms to form W-N bonds. In this step, the formation of W-N bonds will introduce tensile strain to adjacent W-chalcogen bonds given the shorter bond length of W-N (2.12 Å) than the replaced W-chalcogen bonds (2.43 Å ~ 2.74 Å).[14,61] (iii.-iv.) Generation of more $V_x$ at the strained sites, as the tensile strain introduced in step (ii.-iii.) lowers the formation energy of $V_x$.[62] (iv.-v.) Nitrogen atoms occupy $V_x$ sites and bond with adjacent W atoms, consequently introducing more strained sites around the new formed W-N bonds. (v.-vi.) Construction of WN$_x$ crystal through the iteration of step (i.-ii.) to (iv.-v.).

In the proposed reaction procedure, changing the chalcogen elements in W-TMDs precursors has the most prominent impact on step (i.-ii.) since the formation energy of V$_x$ varies significantly with different chalcogen elements.[63] We compare the formation energy of V$_x$ (adapted from Komsa *et al.*'s work)[63] along with the T$_{th}$ of WS$_2$, WSe$_2$ and WTe$_2$ in Figure 5b. A positive correlation is observed where the T$_{th}$ increases with the formation energy of V$_x$. The higher formation energy elevates the temperature required for the generation of V$_x$. Thus, the chalcogen atoms in the TMDs can serve as a useful knob to tune the conversion condition for desired applications.

Apart from theoretical analysis, we also experimentally demonstrate the important role of chalcogen vacancies in initiating the conversion process. We performed conversion on WSe$_2$ flake at a gentle condition (e.g. 700 °C for 10 minutes each run, and iterate for three times). Optical images of the flake at different stages are presented in Figure 5c-f. As shown in Figure 5d, clear sign of reaction proceeding from edge to middle is observed with minimum reaction in the middle of the flake after 10 minutes of reaction. After 20 minutes of the reaction (Figure 5e), the reaction proceeds progressively from edge to the center compared to 10 minutes result. Additionally, circular spots of the converted areas show up on the surface, far from the edges of the flake. Larger portion of the flake is converted in a similar fashion after 30 minutes of reaction. In pristine WSe$_2$ flakes, the surface of the flake contains minimal V$_{Se}$ density while the edge of the flake is rich of V$_{Se}$. Therefore, majority of the conversion happens at the edge of the flake, which is consistent with the proposed defect-initiated process. After 10 minutes of reaction, more V$_{Se}$ are generated on the surface due to thermal activation of Se atoms, creating initiating spots of the reaction on the surface far from the edge of the WSe$_2$ flake. Hence the conversion is also observed in the middle of the surface.

To support the strain-driven conversion mechanism, we performed TEM characterization on an intermediate state of the conversion. The TEM image in Figure 5g provides detailed view of the WSe$_2$-W$_5$N$_6$ lateral heterostructure with a sharp interface. Inset of Figure 5g shows the optical image of the sample. SAED patterns are obtained at different locations of the sample including the unconverted WSe$_2$ region (Figure 5h), the WSe$_2$-W$_5$N$_6$ interface (Figure 5i), and the fully converted W$_5$N$_6$ region (Figure 5j). Moreover, we extracted the interplanar distance between (100) planes (d$_{100}$) of WSe$_2$ from diffraction spots in Figure 5h and 5i, and show the values in Table S1. A larger d$_{100}$ of WSe$_2$ is observed at the WSe$_2$-W$_5$N$_6$ interface (*i.e.*, 2.95 Å) compared to the region containing solely unconverted WSe$_2$ (*i.e.*, 2.93 Å), revealing the WSe$_2$ lattice expansion at the interface. This is attributed to the tensile strain applied onto WSe$_2$ lattice by the formation of W-N bonds, as the bond length of W-N (*i.e.*, 2.12 Å) is shorter than that of W-Se (*i.e.*, 2.43 Å ~ 2.74 Å). [14,61] This result further supports our strain-driven conversion mechanism.

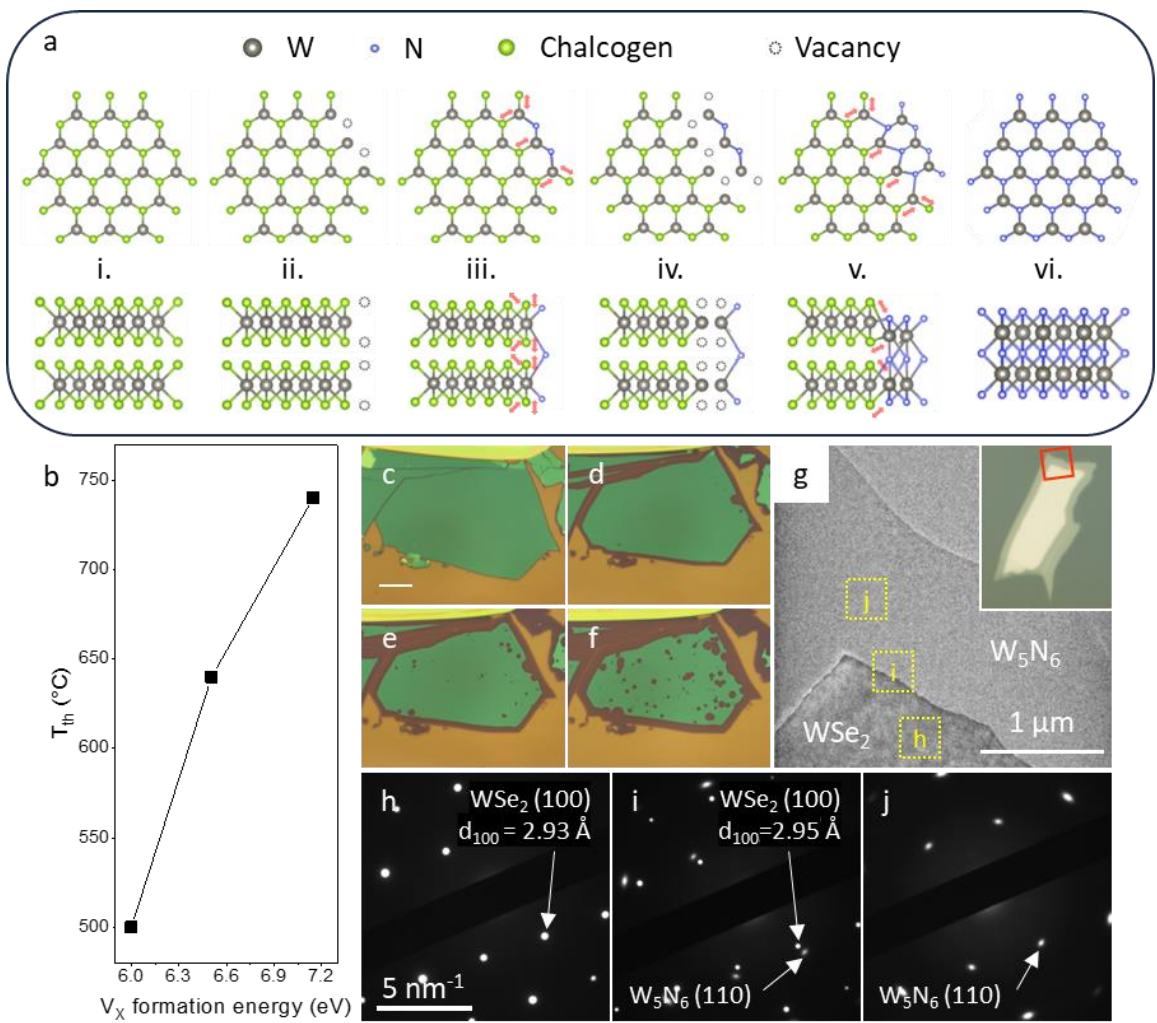

**Figure 5 The role of chalcogen atoms in W-TMDs in the conversion process for WN$_x$.** (a) Schematic illustration of the proposed reaction progress. Top view in the first row and side view in the second row. (i.-ii.) Generation of chalcogen vacancy. (ii.-iii.) Chalcogen vacancy occupied by nitrogen atom and introduce strain to adjacent chalcogen sites. (iii.-iv.) Liberation of strained chalcogen atoms and creating vacancy. (iv.-v.) Chalcogen vacancies occupied by nitrogen atoms resulting in progressive reaction. (v.-vi.) Conversion of entire crystal by iteration of above steps. (b) Correlation between reactivity and the formation energy of chalcogen vacancies in different W-TMDs. y-axis presents the threshold temperature of the nitridation reaction of each W-TMDs. X-axis presents the formation energy of chalcogen vacancies adapted from calculation reported by Komsa *et al.*[63] (c-f) Optical images of WSe$_2$ at intermediate conversion states. The sample is converted at 700 °C with iteration of 10 minutes: (c) pristine, (d) 10 minutes, (e) 20 minutes, (f) 30 minutes. All images share the same scale bar in (c) (10 μm). (g) TEM image of the WSe$_2$-W$_5$N$_6$ heterostructure. Inset shows the optical image of the sample on TEM grid. The TEM image is taken in the squared area labeled in the inset. (h-j) SAED patterns taken at different locations: (h) unconverted WSe$_2$, (i) WSe$_2$-W$_5$N$_6$ interface, and (j) converted W$_5$N$_6$. Each location is labeled in (g).

**Conclusion**

We report the downscaling of non vdW $W_5N_6$ material with tunable thicknesses in the range from 2.9 nm to 45.6 nm through an atomic substitution reaction from $WSe_2$ crystals. The obtained ultrathin $W_5N_6$ flakes exhibit high crystallinity and smooth surfaces, in contrast to the rough surface of the thin films of other non vdW materials such as Cu at similar thicknesses. From electrical measurements, we show that the conductivity of 2D $W_5N_6$ persists from 45.6 nm to 2.9 nm, in contrast to the significant resistivity scaling of vacuum deposited Cu thin films in a similar thickness region. We also observed a weak gate tunability of the conductivity of 2D $W_5N_6$, attributed to the semimetallic band structure of the material. Additionally, we propose a chalcogen vacancy-initiated and strain-driven substitution reaction, highlighting the role of chalcogen anion in controlling the reactivity and the conversion temperature. Thus, various W-TMDs precursors with different chalcogen anion could offer a useful knob to tune the downscaling process, enhancing the potential for materials manufacturing and integration.

**Methods**

*Synthesis of 2D $WN_x$:* $WS_2$, $WSe_2$, and $WTe_2$ flakes of various thicknesses were mechanically exfoliated from their bulk crystals using a scotch tape, and then transferred onto a $SiO_2/Si$ substrate with a $SiO_2$ layer of 300 nm thick. The substrates were cleaned using similar method in our previous work.[35,64] The conversion is conducted in a quartz tube with Argon (50 sccm) as the carrier gas, heated by a tube furnace. A flask containing ammonium hydroxide aqua solution (30 wt%, Thermo Fisher) is mounted at the upper stream of gas line to generate ammonia gas through evaporation, and the $SiO_2/Si$ chip with W-TMDs flakes on it is placed in the middle of the heating zone. The temperature is set to ramp from room temperature to the target temperature (500 ~ 950 °C) under the rate of 50 °C/min, followed by maintaining at the target temperature for 30 minutes for the conversion. Afterwards, the furnace is cooled to room temperature rapidly by a fan with lid of the furnace opened.

*Materials characterizations:* As-prepared $W_5N_6$ were characterized using AFM, Raman spectroscopy, XPS, and high-resolution TEM. The AFM topography was acquired using a Bruker Dimension system. Raman measurements were performed on a Renishaw inVia Raman microscope equipped with a 532 nm laser line. All spectra were taken using the same condition with a laser power of 1.6 mW, beam diameter of 1 µm, and acquisition time of 60 s. The XPS is carried out on Nexsa G2, from Thermo Fisher Scientific. The spot size of the X-ray that shined on the material surface is 100 µm in diameter in the measurements. Planar TEM measurements were performed using a FEI Tecnai Osiris transmission electron microscope, operating at a 200-kV accelerating voltage.

For the cross-sectional STEM measurements, the electron beam transparent specimen was prepared using the Thermo Scientific Scios 2 DualBeam focused ion beam scanning electron microscopy, and the images were taken using FEI Titan3 G2 aberration-corrected TEM/STEM, operating at a 300-kV accelerating voltage with a high-angle annular dark-field (HAADF) detector.

*Electrode fabrication and electrical measurements:* All the electrodes were fabricated inside a cleanroom-in-a-glovebox.[44] The electrode patterning was done using a bilayer photoresist (LOR1A/S1805) and with a maskless laser lithography system (Heidelberg Instruments). Then the Cr/Au (5 nm/45 nm) were deposited using e-beam evaporation (Angstrom Engineering) followed by lift-off using Remover PG (Kayaku). I-V characteristic curves are measured using a Keithley 2400 source meter under ambient conditions (room temperature in air).

*Fabrication of $MoS_2$ transistors with Au and $W_5N_6$ contacts:* $MoS_2$ with Au contact is fabricated by laser lithography followed by metal deposition, similar to the description in the *Electrode fabrication and electrical measurements* section. $MoS_2$ with $W_5N_6$ contacts are fabricated by the following steps, as shown in Figure S12: (1) Etch $WSe_2$ precursor flakes into stripes with gaps in the middle. (2) Convert etched $WSe_2$ into $W_5N_6$. (3) Fabricate Ti/Au electrodes on $W_5N_6$ by lithography and deposition. (4) Fabricate insulating $Al_2O_3$ layer on the Au electrode to avoid direct contact between Au and $MoS_2$. (5) Transfer and stack a monolayer $MoS_2$ flake onto $W_5N_6$ to form the $W_5N_6$-$MoS_2$ vdW heterostructure. The stacked sample is then annealed in vacuum (30 mTorr, 200 °C, 30 minutes) to improve adhesion of the flake.

*Binding energy calculations:* The ground-state total energy of the AA- and AB-$W_5N_6$ are calculated using the density functional theory with the Quantum ESPRESSO package.[65] We use a cut-off energy of about 60 Ry for the plane wave and a k-point mesh of $8 \times 8 \times 8$ for Brillouin zone integration. The valence electron-ion core interactions and the exchange-correlation functional are adopted by the projector augmented wave potentials (PAW) and Perdew–Burke–Ernzerhof (PBE) approximation,[66,67] respectively. The atomic structures are optimized by the Broyden-Fretcher-Goldfarb–Shanno algorithm,[68] in which convergence criteria are $1 \times 10^{-5}$ Ry/a.u. for atomic forces and $5 \times 10^{-2}$ GPa for all stress components. The binding energy is given by $E_b(S) = E_{total}(S) - 5E(W) - 6E(N)$, where $E_{total}(S)$ are calculated ground-state total energies of S = AA-$W_5N_6$ or AB-$W_5N_6$ per unit cell, and $E(W)$ and $E(N)$ are the total energies of a W atom in the body-centered-cubic crystal and a N atom in the $N_2$ molecule, respectively.

*Phonon calculations:* The phonon dispersions of AA- and AB-$W_5N_6$ are calculated using the density-functional perturbation theory (DFPT).[69] The q-point mesh of $4 \times 4 \times 4$ is selected based on the phonon frequency convergence.

**Supporting Information**

The Supporting Information is available free of charge online. Supporting information presents supplementary data of $W_5N_6$ phonon calculations, Raman spectra, XPS survey, TEM images, simulated SAED patterns, $W_5N_6/MoS_2$ FET measurements, surface roughness measurements, and optical images of conversion of different W-TMDs. A table in supporting information also summarizes the lattice constant of $WSe_2$ in the unconverted $WSe_2$ region and at the $WSe_2$-$W_5N_6$ interface.

**Acknowledgment**

This material is based upon work supported by the U.S. Department of Energy (DOE), Office of Science, Basic Energy Science (BES) under Award Number DE-SC0021064 (X.L. and H.G.) and DE-SC0018675 (K.S.B. and M.G.). X.L. acknowledges the membership of the Photonics Center at Boston University. H. G. acknowledges the support of BUnano fellowship from Boston University Nanotechnology Innovation Center. Work done by X.L. is also supported by the National Science Foundation (NSF) under Grant No. 1945364 and No. 2216008. D. Z. and M. T. would like to acknowledge the financial support from AFOSR (FA9550-23-1-0447) and the Basic Office of Science of the Department of Energy (Award No. DE-SC0018025). N.T.H. acknowledges the Researcher, Young Leaders Overseas Program from Tohoku University. R.S. acknowledges a JSPS KAKENHI Grant (No. JP22H00283) and the Yushan Fellow Program by the Ministry of Education (MOE), Taiwan. G.N. is grateful for the support of the National Science Foundation (NSF) EPMD program via grant 2211334.

**Conflict of Interest**

The authors declare no conflict of interest.

**References**

(1) Ningthoujam, R. S.; Gajbhiye, N. S. Synthesis, Electron Transport Properties of Transition Metal Nitrides and Applications. *Prog. Mater. Sci.* **2015**, *70*, 50–154.

(2) Lévy, F.; Hones, P.; Schmid, P. E.; Sanjinés, R.; Diserens, M.; Wiemer, C. Electronic States and Mechanical Properties in Transition Metal Nitrides. *Surf. Coatings Technol.* **1999**, *120–121*, 284–290.

(3) Barnett, S. A.; Madan, A. Hardness and Stability of Metal-Nitride Nanoscale Multilayers. *Scr. Mater.* **2004**, *50* (6), 739–744.

(4) Kindlund, H.; Sangiovanni, D. G.; Petrov, I.; Greene, J. E.; Hultman, L. A Review of the Intrinsic Ductility and Toughness of Hard Transition-Metal Nitride Alloy Thin Films. *Thin Solid Films* **2019**, *688*, 137479.

(5) Salamat, A.; Hector, A. L.; Kroll, P.; McMillan, P. F. Nitrogen-Rich Transition Metal Nitrides. *Coord. Chem. Rev.* **2013**, *257* (13–14), 2063–2072.

(6) Wang, H.; Li, J.; Li, K.; Lin, Y.; Chen, J.; Gao, L.; Nicolosi, V.; Xiao, X.; Lee, J. M. Transition Metal Nitrides for Electrochemical Energy Applications. *Chem. Soc. Rev.* **2021**, *50* (2), 1354–1390.

(7) Shi, J.; Jiang, B.; Li, C.; Yan, F.; Wang, D.; Yang, C.; Wan, J. Review of Transition Metal Nitrides and Transition Metal Nitrides/Carbon Nanocomposites for Supercapacitor Electrodes. *Mater. Chem. Phys.* **2020**, *245*, 122533.

(8) Zhou, Y.; Guo, W.; Li, T. A Review on Transition Metal Nitrides as Electrode Materials for Supercapacitors. *Ceram. Int.* **2019**, *45* (17), 21062–21076.

(9) Wittmer, M. Interfacial Reactions between Aluminum and Transition-Metal Nitride and Carbide Films. *J. Appl. Phys.* **1982**, *53* (2), 1007–1012.

(10) Kim, H.; Cabral, C.; Lavoie, C.; Rossnagel, S. M. Diffusion Barrier Properties of Transition Metal Thin Films Grown by Plasma-Enhanced Atomic-Layer Deposition. *J. Vac. Sci. Technol. B Microelectron. Nanom. Struct. Process. Meas. Phenom.* **2002**, *20* (4), 1321–1326.

(11) Allen, M. J.; Tung, V. C.; Kaner, R. B. Honeycomb Carbon: A Review of Graphene. *Chem. Rev.* **2010**, *110* (1), 132–145.

(12) Samy, O.; Zeng, S.; Birowosuto, M. D.; El Moutaouakil, A. A Review on $MoS_2$ Properties, Synthesis, Sensing Applications and Challenges. *Crystals* **2021**, *11* (4), 1–24.

(13) Jauberteau, I.; Bessaudou, A.; Mayet, R.; Cornette, J.; Jauberteau, J. L.; Carles, P.; Merle-Méjean, T. Molybdenum Nitride Films: Crystal Structures, Synthesis, Mechanical, Electrical and Some Other Properties. *Coatings*. **2015**, pp 656–687.

(14) Zhao, Z.; Bao, K.; Duan, D.; Tian, F.; Huang, Y.; Yu, H.; Liu, Y.; Liu, B.; Cui, T. The Low Coordination Number of Nitrogen in Hard Tungsten Nitrides: A First-Principles Study. *Phys. Chem. Chem. Phys.* **2015**, *17* (20), 13397–13402.

(15) Patsalas, P.; Kalfagiannis, N.; Kassavetis, S.; Abadias, G.; Bellas, D. V.; Lekka, C.; Lidorikis, E. Conductive Nitrides: Growth Principles, Optical and Electronic Properties, and Their Perspectives in Photonics and Plasmonics. *Mater. Sci. Eng. R Reports* **2018**, *123*, 1–55.

(16) Lozovoy, K. A.; Korotaev, A. G.; Kokhanenko, A. P.; Dirko, V. V.; Voitsekhovskii, A. V. Kinetics of Epitaxial Formation of Nanostructures by Frank–van Der Merwe, Volmer–Weber and Stranski–Krastanow Growth Modes. *Surf. Coatings Technol.* **2020**, *384*, 125289.

(17) Gao, H.; Wang, Z.; Cao, J.; Lin, Y. C.; Ling, X. Advancing Nanoelectronics Applications: Progress in Non-van Der Waals 2D Materials. *ACS Nano* **2024**, *18* (26), 16343–16358.

(18) Chen, J.; Gao, J. Strong Electron–Phonon Coupling in 3D WN and Coexistence of Intrinsic Superconductivity and Topological Nodal Line in Its 2D Limit. *Phys. Status Solidi - Rapid Res. Lett.* **2022**, *16* (1), 1–10.

(19) Campi, D.; Kumari, S.; Marzari, N. Prediction of Phonon-Mediated Superconductivity with High Critical Temperature in the Two-Dimensional Topological Semimetal $W_2N_3$. *Nano Lett.* **2021**, *21* (8), 3435–3442.

(20) Chin, H.-T.; Wang, D.-C.; Gulo, D. P.; Yao, Y.-C.; Yeh, H.-C.; Muthu, J.; Chen, D.-R.; Kao, T.-C.; Kalbáč, M.; Lin, P.-H.; Cheng, C.-M.; Hofmann, M.; Liang, C.-T.; Liu, H.-L.; Chuang, F.-C.; Hsieh, Y.-P. Tungsten Nitride ($W_5N_6$): An Ultraresilient 2D Semimetal. *Nano Lett.* **2023**, *24* (1), 67–73.

(21) Shen, P. C.; Su, C.; Lin, Y.; Chou, A. S.; Cheng, C. C.; Park, J. H.; Chiu, M. H.; Lu, A. Y.; Tang, H. L.; Tavakoli, M. M.; Pitner, G.; Ji, X.; Cai, Z.; Mao, N.; Wang, J.; Tung, V.; Li, J.; Bokor, J.; Zettl, A.; Wu, C. I.; Palacios, T.; Li, L. J.; Kong, J. Ultralow Contact Resistance between Semimetal and Monolayer Semiconductors. *Nature* **2021**, *593* (7858), 211–217.

(22) Plisson, V. M.; Yao, X.; Wang, Y.; Varnavides, G.; Suslov, A.; Graf, D.; Choi, E. S.; Yang, H. Y.; Wang, Y.; Romanelli, M.; McNamara, G.; Singh, B.; McCandless, G. T.; Chan, J. Y.; Narang, P.; Tafti, F.; Burch, K. S. Engineering Anomalously Large Electron Transport in Topological Semimetals. *Adv. Mater.* **2024**, *36* (24), 1–11.

(23) Han, H. J.; Kumar, S.; Jin, G.; Ji, X.; Hart, J. L.; Hynek, D. J.; Sam, Q. P.; Hasse, V.; Felser, C.; Cahill, D. G.; Sundararaman, R.; Cha, J. J. Topological Metal MoP Nanowire for Interconnect. *Adv. Mater.* **2023**, *35* (13), 1–8.

(24) Yu, H.; Yang, X.; Xiao, X.; Chen, M.; Zhang, Q.; Huang, L.; Wu, J.; Li, T.; Chen, S.; Song, L.; Gu, L.; Xia, B. Y.; Feng, G.; Li, J.; Zhou, J. Atmospheric-Pressure Synthesis of 2D Nitrogen-Rich Tungsten Nitride. *Adv. Mater.* **2018**, *30* (51), 1–7.

(25) Xiao, X.; Yu, H.; Jin, H.; Wu, M.; Fang, Y.; Sun, J.; Hu, Z.; Li, T.; Wu, J.; Huang, L.; Gogotsi, Y.; Zhou, J. Salt-Templated Synthesis of 2D Metallic MoN and Other Nitrides. *ACS Nano* **2017**, *11* (2), 2180–2186.

(26) Kong, Q.; Liu, D.; Yang, L.; Zhao, H.; Zhang, J.; Xi, G. Tungsten Nitride with a Two-Dimensional Multilayer Structure for Boosting the Surface-Enhanced Raman Effect. *J. Phys. Chem. Lett.* **2023**, 10894–10899.

(27) Chin, H. T.; Wang, D. C.; Wang, H.; Muthu, J.; Khurshid, F.; Chen, D. R.; Hofmann, M.; Chuang, F. C.; Hsieh, Y. P. Confined VLS Growth of Single-Layer 2D Tungsten Nitrides. *ACS Appl. Mater. Interfaces* **2024**, *16* (1), 1705–1711.

(28) Wang, H.; Sandoz-Rosado, E. J.; Tsang, S. H.; Lin, J.; Zhu, M.; Mallick, G.; Liu, Z.; Teo, E. H. T. Elastic Properties of 2D Ultrathin Tungsten Nitride Crystals Grown by Chemical Vapor Deposition. *Adv. Funct. Mater.* **2019**, *29* (31), 1–7.

(29) Zhang, J.; Chen, J.; Luo, Y.; Chen, Y.; Wei, X.; Wang, G.; Wang, R. Sandwich-like Electrode with Tungsten Nitride Nanosheets Decorated with Carbon Dots as Efficient Electrocatalyst for Oxygen Reduction. *Appl. Surf. Sci.* **2019**, *466* (July 2018), 911–919.

(30) Villaseca, L.; Moreno, B.; Lorite, I.; Jurado, J. R.; Chinarro, E. Synthesis and Characterization of Tungsten Nitride ($W_2N$) from $WO_3$ and $H_2WO_4$ to Be Used in the Electrode of Electrochemical Devices. *Ceram. Int.* **2015**, *41* (3), 4282–4288.

(31) Schmiedl, E.; Wissmann, P.; Finzel, H. U. The Electrical Resistivity of Ultra-Thin Copper Films. *Zeitschrift fur Naturforsch. - Sect. A* **2008**, *63* (10–11), 739–744.

(32) Gall, D. The Search for the Most Conductive Metal for Narrow Interconnect Lines. *J. Appl. Phys.* **2020**, *127* (5).

(33) Cao, J.; Li, T.; Gao, H.; Lin, Y.; Wang, X.; Wang, H.; Palacios, T.; Ling, X. Realization of 2D Crystalline Metal Nitrides via Selective Atomic Substitution. *Sci. Adv.* **2020**, *6* (2), eaax8784.

(34) Cao, J.; Li, T.; Gao, H.; Cong, X.; Lin, M. L.; Russo, N.; Luo, W.; Ding, S.; Wang, Z.; Smith, K. E.; Tan, P. H.; Ma, Q.; Ling, X. Ultrathin GaN Crystal Realized Through Nitrogen Substitution of Layered GaS. *J. Electron. Mater.* **2023**, *52* (11), 7554–7565.

(35) Gao, H.; Cao, J.; Li, T.; Luo, W.; Gray, M.; Kumar, N.; Burch, K. S.; Ling, X. Phase-Controllable Synthesis of Ultrathin Molybdenum Nitride Crystals Via Atomic Substitution of $MoS_2$. *Chem. Mater.* **2022**, *34* (1), 351–357.

(36) Zeng, H.; Liu, G. Bin; Dai, J.; Yan, Y.; Zhu, B.; He, R.; Xie, L.; Xu, S.; Chen, X.; Yao, W.; Cui, X. Optical Signature of Symmetry Variations and Spin-Valley Coupling in Atomically Thin Tungsten Dichalcogenides. *Sci. Rep.* **2013**, *3*, 2–6.

(37) Xing, W.; Miao, X.; Meng, F.; Yu, R. Crystal Structure of and Displacive Phase Transition in Tungsten Nitride WN. *J. Alloys Compd.* **2017**, *722* (193), 517–524.

(38) Yang, L.; Majumdar, K.; Liu, H.; Du, Y.; Wu, H.; Hatzistergos, M.; Hung, P. Y.; Tieckelmann, R.; Tsai, W.; Hobbs, C.; Ye, P. D. Chloride Molecular Doping Technique on 2D Materials: $WS_2$ and $MoS_2$. *Nano Lett.* **2014**, *14* (11), 6275–6280.

(39) Fujiwara, K.; Tsukazaki, A. Formation of Distorted Rutile-Type $NbO_2$, $MoO_2$, and

WO$_2$ Films by Reactive Sputtering. *J. Appl. Phys.* **2019**, *125* (8).

(40) Huang, J.; Yang, L.; Liu, D.; Chen, J.; Fu, Q.; Xiong, Y.; Lin, F.; Xiang, B. Large-Area Synthesis of Monolayer WSe$_2$ on a SiO$_2$/Si Substrate and Its Device Applications. *Nanoscale* **2015**, *7* (9), 4193–4198.

(41) Sowa, M. J.; Yemane, Y.; Prinz, F. B.; Provine, J. Plasma-Enhanced Atomic Layer Deposition of Tungsten Nitride. *J. Vac. Sci. Technol. A Vacuum, Surfaces, Film.* **2016**, *34* (5), 051516.

(42) Galtayries, A.; Laksono, E.; Siffre, J. M.; Argile, C.; Marcus, P. XPS Study of the Adsorption of NH$_3$ on Nickel Oxide on Ni(111). *Surf. Interface Anal.* **2000**, *30* (1), 140–144.

(43) Sreedhara, M. B.; Miroshnikov, Y.; Zheng, K.; Houben, L.; Hettler, S.; Arenal, R.; Pinkas, I.; Sinha, S. S.; Castelli, I. E.; Tenne, R. Nanotubes from Ternary WS$_{2(1-x)}$Se$_{2x}$ Alloys: Stoichiometry Modulated Tunable Optical Properties. *J. Am. Chem. Soc.* **2022**, *144* (23), 10530–10542.

(44) Gray, M. J.; Kumar, N.; O'Connor, R.; Hoek, M.; Sheridan, E.; Doyle, M. C.; Romanelli, M. L.; Osterhoudt, G. B.; Wang, Y.; Plisson, V.; Lei, S.; Zhong, R.; Rachmilowitz, B.; Zhao, H.; Kitadai, H.; Shepard, S.; Schoop, L. M.; Gu, G. D.; Zeljkovic, I.; Ling, X.; Burch, K. S. A Cleanroom in a Glovebox. *Rev. Sci. Instrum.* **2020**, *91* (7).

(45) Borah, A.; Nipane, A.; Choi, M. S.; Hone, J.; Teherani, J. T. Low-Resistance p-Type Ohmic Contacts to Ultrathin WSe$_2$ by Using a Monolayer Dopant. *ACS Appl. Electron. Mater.* **2021**, *3* (7), 2941–2947.

(46) Sui, Y.; Appenzeller, J. Screening and Interlayer Coupling in Multilayer Graphene Field-Effect Transistors. *Nano Lett.* **2009**, *9* (8), 2973–2977.

(47) Das, S.; Appenzeller, J. Screening and Interlayer Coupling in Multilayer MoS$_2$. *Phys. Status Solidi - Rapid Res. Lett.* **2013**, *7* (4), 268–273.

(48) Dulmaa, A.; Cougnon, F. G.; Dedoncker, R.; Depla, D. On the Grain Size-Thickness Correlation for Thin Films. *Acta Mater.* **2021**, *212*, 116896.

(49) Park, Y. B.; Jeong, C.; Guo, L. J. Resistivity Scaling Transition in Ultrathin Metal Film at Critical Thickness and Its Implication for the Transparent Conductor Applications. *Adv. Electron. Mater.* **2022**, *8* (3), 1–7.

(50) Wang, W.; Qi, J.; Zhai, L.; Ma, C.; Ke, C.; Zhai, W.; Wu, Z.; Bao, K.; Yao, Y.; Li, S.; Chen, B.; Repaka, D. V. M.; Zhang, X.; Ye, R.; Lai, Z.; Luo, G.; Chen, Y.; He, Q. Preparation of 2D Molybdenum Phosphide via Surface-Confined Atomic Substitution. *Adv. Mater.* **2022**, *34* (35), 1–11.

(51) Foadi, F.; Ten Brink, G. H.; Mohammadizadeh, M. R.; Palasantzas, G. Roughness Dependent Wettability of Sputtered Copper Thin Films: The Effect of the Local Surface Slope. *J. Appl. Phys.* **2019**, *125* (24), 244307.

(52) Jiang, Y. C.; Gao, J.; Wang, L. Raman Fingerprint for Semi-Metal WTe$_2$ Evolving

from Bulk to Monolayer. *Sci. Rep.* **2016**, *6*, 1–7.

(53) Jeon, J.; Park, Y.; Choi, S.; Lee, J.; Lim, S. S.; Lee, B. H.; Song, Y. J.; Cho, J. H.; Jang, Y. H.; Lee, S. Epitaxial Synthesis of Molybdenum Carbide and Formation of a Mo2C/MoS$_2$ Hybrid Structure via Chemical Conversion of Molybdenum Disulfide. *ACS Nano* **2018**, *12* (1), 338–346.

(54) Sreepal, V.; Yagmurcukardes, M.; Vasu, K. S.; Kelly, D. J.; Taylor, S. F. R.; Kravets, V. G.; Kudrynskyi, Z.; Kovalyuk, Z. D.; Patanè, A.; Grigorenko, A. N.; Haigh, S. J.; Hardacre, C.; Eaves, L.; Sahin, H.; Geim, A. K.; Peeters, F. M.; Nair, R. R. Two-Dimensional Covalent Crystals by Chemical Conversion of Thin van Der Waals Materials. *Nano Lett.* **2019**, *19* (9), 6475–6481.

(55) Zhao, M.; Yang, S.; Zhang, K.; Zhang, L.; Chen, P.; Yang, S.; Zhao, Y.; Ding, X.; Zu, X.; Li, Y.; Zhao, Y.; Qiao, L.; Zhai, T. A Universal Atomic Substitution Conversion Strategy Towards Synthesis of Large-Size Ultrathin Nonlayered Two-Dimensional Materials. *Nano-Micro Lett.* **2021**, *13* (1), 1–13.

(56) Lam, D.; Lebedev, D.; Hersam, M. C. Morphotaxy of Layered van Der Waals Materials. *ACS Nano* **2022**, *16* (5), 7144–7167.

(57) Lai, S.; Byeon, S.; Jang, S. K.; Lee, J.; Lee, B. H.; Park, J. H.; Kim, Y. H.; Lee, S. HfO$_2$/HfS$_2$ Hybrid Heterostructure Fabricated: Via Controllable Chemical Conversion of Two-Dimensional HfS$_2$. *Nanoscale* **2018**, *10* (39), 18758–18766.

(58) Chamlagain, B.; Cui, Q.; Paudel, S.; Cheng, M. M. C.; Chen, P. Y.; Zhou, Z. Thermally Oxidized 2D TaS$_2$ as a High-κ Gate Dielectric for MoS$_2$ Field-Effect Transistors. *2D Mater.* **2017**, *4* (3), 031002.

(59) Li, T.; Cao, J.; Gao, H.; Wang, Z.; Geiwitz, M.; Burch, K. S.; Ling, X. Epitaxial Atomic Substitution for MoS$_2$-MoN Heterostructure Synthesis. *ACS Appl. Mater. Interfaces* **2022**, *14* (51), 57144–57152.

(60) Jin, T.; Zheng, Y.; Gao, J.; Wang, Y.; Li, E.; Chen, H.; Pan, X.; Lin, M.; Chen, W. Controlling Native Oxidation of HfS$_2$ for 2D Materials Based Flash Memory and Artificial Synapse. *ACS Appl. Mater. Interfaces* **2021**, *13* (8), 10639–10649.

(61) Amin, B.; Kaloni, T. P.; Schwingenschlögl, U. Strain Engineering of WS$_2$, WSe$_2$, and WTe$_2$. *RSC Adv.* **2014**, *4* (65), 34561–34565.

(62) Aschauer, U.; Pfenninger, R.; Selbach, S. M.; Grande, T.; Spaldin, N. A. Strain-Controlled Oxygen Vacancy Formation and Ordering in CaMnO$_3$. *Phys. Rev. B - Condens. Matter Mater. Phys.* **2013**, *88* (5), 1–7.

(63) Komsa, H. P.; Kotakoski, J.; Kurasch, S.; Lehtinen, O.; Kaiser, U.; Krasheninnikov, A. V. Two-Dimensional Transition Metal Dichalcogenides under Electron Irradiation: Defect Production and Doping. *Phys. Rev. Lett.* **2012**, *109* (3), 1–5.

(64) Yao, C. H.; Gao, H.; Ping, L.; Gulo, D. P.; Liu, H. L.; Tuan Hung, N.; Saito, R.; Ling, X. Nontrivial Raman Characteristics in 2D Non-Van Der Waals Mo$_5$N$_6$. *ACS Nano* **2024**.

(65) Giannozzi, P.; Baroni, S.; Bonini, N.; Calandra, M.; Car, R.; Cavazzoni, C.; Ceresoli, D.; Chiarotti, G. L.; Cococcioni, M.; Dabo, I.; Dal Corso, A.; De Gironcoli, S.; Fabris, S.; Fratesi, G.; Gebauer, R.; Gerstmann, U.; Gougoussis, C.; Kokalj, A.; Lazzeri, M.; Martin-Samos, L.; Marzari, N.; Mauri, F.; Mazzarello, R.; Paolini, S.; Pasquarello, A.; Paulatto, L.; Sbraccia, C.; Scandolo, S.; Sclauzero, G.; Seitsonen, A. P.; Smogunov, A.; Umari, P.; Wentzcovitch, R. M. QUANTUM ESPRESSO: A Modular and Open-Source Software Project for Quantum Simulations of Materials. *J. Phys. Condens. Matter* **2009**, *21* (39).

(66) Aykol, M.; Kim, S.; Wolverton, C. Van Der Waals Interactions in Layered Lithium Cobalt Oxides. *J. Phys. Chem. C* **2015**, *119* (33), 19053–19058.

(67) Kresse, G., Joubert, D. From Ultrasoft Pseudopotentials to the Projector Augmented-Wave Method. *Phys. Rev. B - Condens. Matter Mater. Phys.* **1999**, *59* (3), 1758–1775.

(68) Hung, N. T.; Nugraha, A. R. T.; Saito, R. *Quantum ESPRESSO Course for Solid-State Physics*; Jenny Stanford Publishing, 2022.

(69) S. Baroni, S. De Gironcoli, A. Dal Corso, P. Giannozzi. Phonons and Related Crystal Properties from Density-Functional Perturbation Theory. *Rev. Mod. Phys.* **2001**, *73* (2), 515.

**Supplementary Materials**

**Table of content**

**Table S1** Lattice parameters extracted from SAED patterns taken at different locations of the WSe$_2$-W$_5$N$_6$ heterostructure.

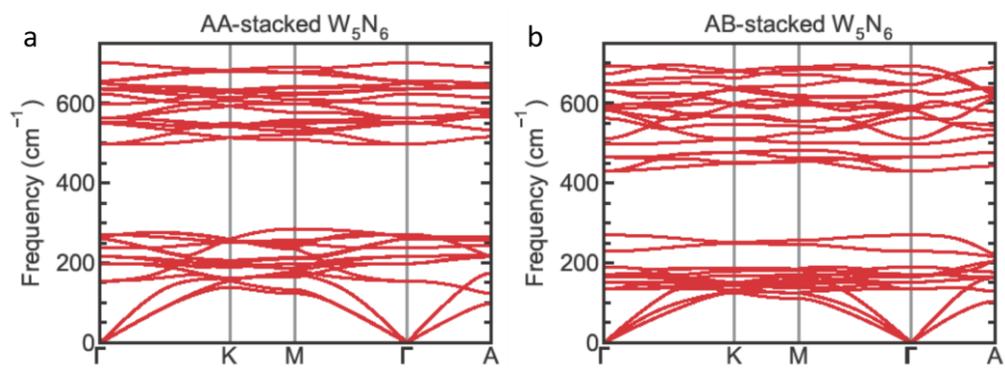

**Figure S1** Phonon dispersion calculation of $W_5N_6$ with different stacking orders. (a) AA stacking of W and N. (b) AB stacking of W and N.

**Evolution of WN$_x$ Raman spectra**

As discussed in the spectroscopy characterization section, we performed the nitridation reaction on WS$_2$, WSe$_2$ and WTe$_2$. To distinguish the conversion on different TMDs, we label the converted flakes as WN$_x$-S, WN$_x$-Se, and WN$_x$-Te for WS$_2$, WSe$_2$, and WTe$_2$ precursors, respectively. The Raman spectra of reaction from different TMDs precursors are shown in Figure S2, and the reaction duration is 30 minutes in each experiment. Three characteristic Raman signals are observed in WN$_x$ converted from different precursors: peaks centred at 198 cm$^{-1}$ (phase 3), 228 cm$^{-1}$ (phase 1), and 256 cm$^{-1}$ (phase 2, *i.e.*, W$_5$N$_6$) are observed with variable intensities under different conditions. The phase order is assigned based on the transition temperature. Namely, phase 1 emerges at lowest temperature, and transit to phase 2 and 3 at elevated temperatures. Note that in most spectra (except for WNx-Te at 600°C), the Raman signatures of more than one phase of WNx are observed, suggesting the phase transition is not complete.

We will focus on WN$_x$-Te for the phase transition since it allows the widest conversion temperature range. In WN$_x$-Te, the 228 cm$^{-1}$ peak is the only feature observed below 600 °C and labeled as phase 1. As conversion temperature raises to 640 °C, a shoulder emerges around 256 cm$^{-1}$ and dominates the spectra once temperature exceeds 750 °C. This suggests a phase transition from phase 1 to W$_5$N$_6$. With temperature keeps increasing, the Raman signal associated to phase 3 emerges at around 198 cm$^{-1}$. For WN$_x$-Se, the converted flakes started with W$_5$N$_6$ dominating the spectra with a shoulder around 230 cm$^{-1}$ suggesting the coexistence of phase 1 and W$_5$N$_6$ at 700 °C. As temperature raises, W$_5$N$_6$ becomes the only phase presented (740 and 800 °C) and eventually taken over by phase 3 (900 °C). Coexistence of multiple phases is always observed from Raman. Such phase transition is attributed to the structural change between isomers of $WN_x$ given their close formation energies.[1,2]

It is worth noting that only $W_5N_6$ have been successfully synthesized with high phase purity and crystallinity using $WSe_2$ as precursor. Although phase 1 can be synthesized using $WTe_2$ precursors with high phase purity, the converted flakes exhibit amorphous crystallinity and hence challenging to identify its crystal structure. As for phase 3, there is always residue from phase 3 and $W_5N_6$. It is also challenging to probe phase 3 without interference from other phases. Therefore, we focus on $W_5N_6$ converted from $WSe_2$ precursors in this work.

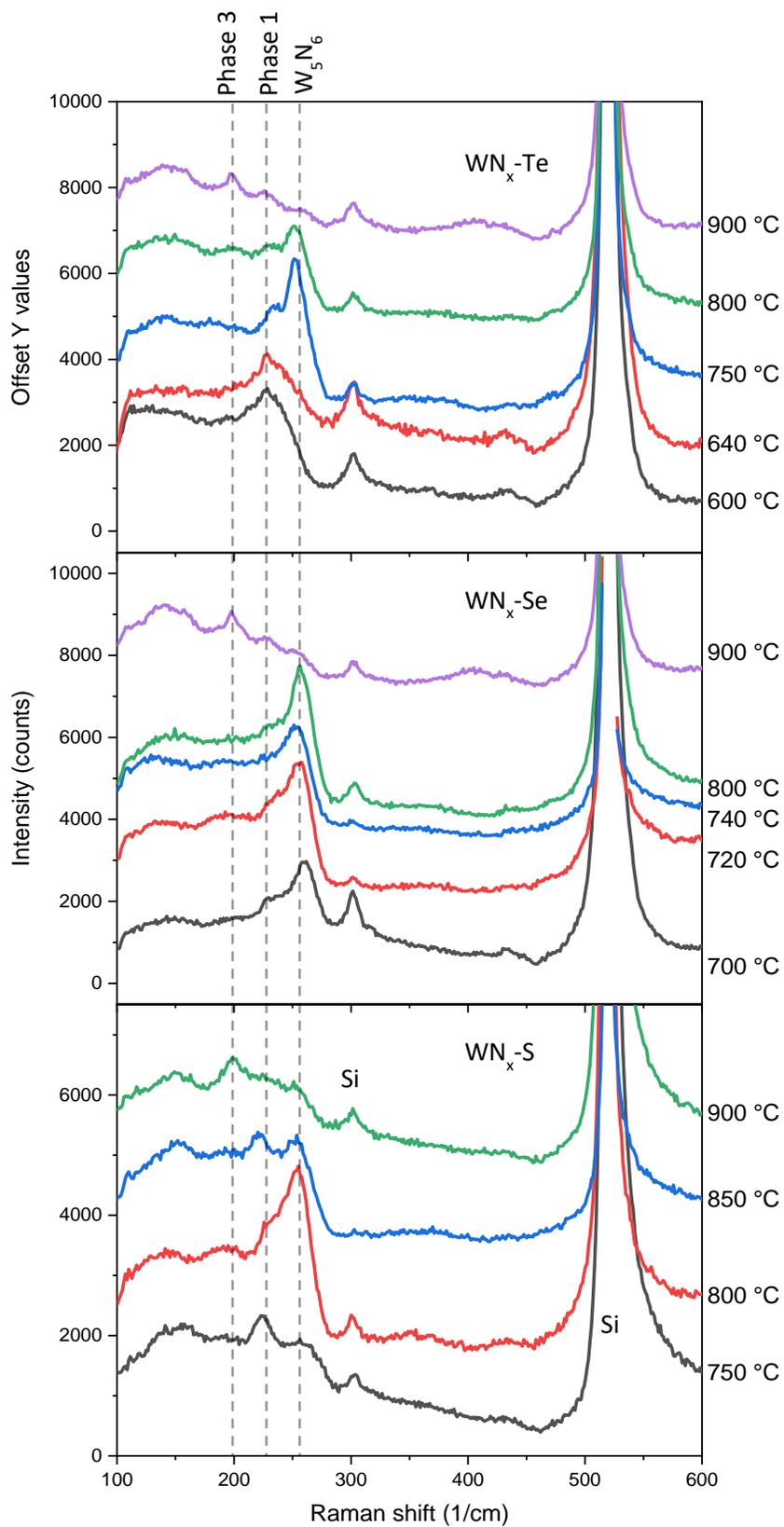

**Figure S2** Raman spectra of WN$_X$-Te, WN$_X$-Se, and WN$_X$-S converted at different temperatures for 30 minutes.

## XPS characterization of W₅N₆

High resolution XPS scans of W and N orbitals are presented in the main text. Figure S2 shows an XPS survey of the W$_5$N$_6$ samples, where the Si and O signals are attributed to the SiO$_2$/Si substrate, C signals are attributed to the measurement chamber, and W and N signals are attributed to the converted flakes. Note that we did not observe any signals from Se (indicated by the dotted line), confirming a complete substitution reaction.

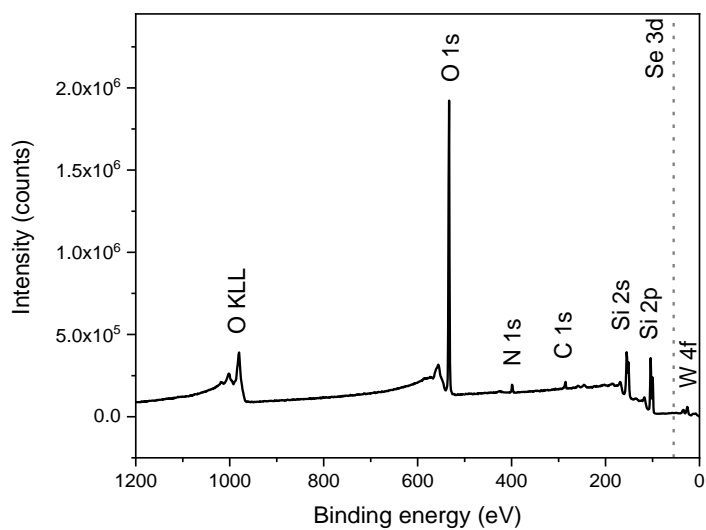

**Figure S3** XPS survey of W$_5$N$_6$. The binding energy of Se 3d orbital is labeled by a dashed line to show the complete removal of Se in the sample.

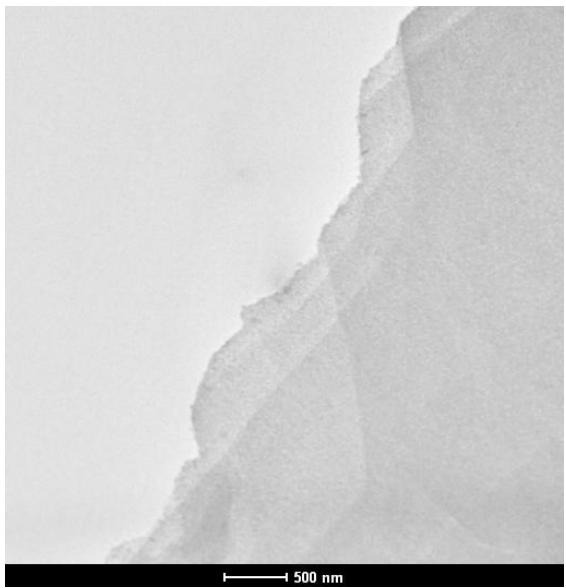

**Figure S4** Low-magnification TEM image of $W_5N_6$. The contrast difference of the image suggests different thickness of the flake, which inherits from the $WSe_2$ precursor.

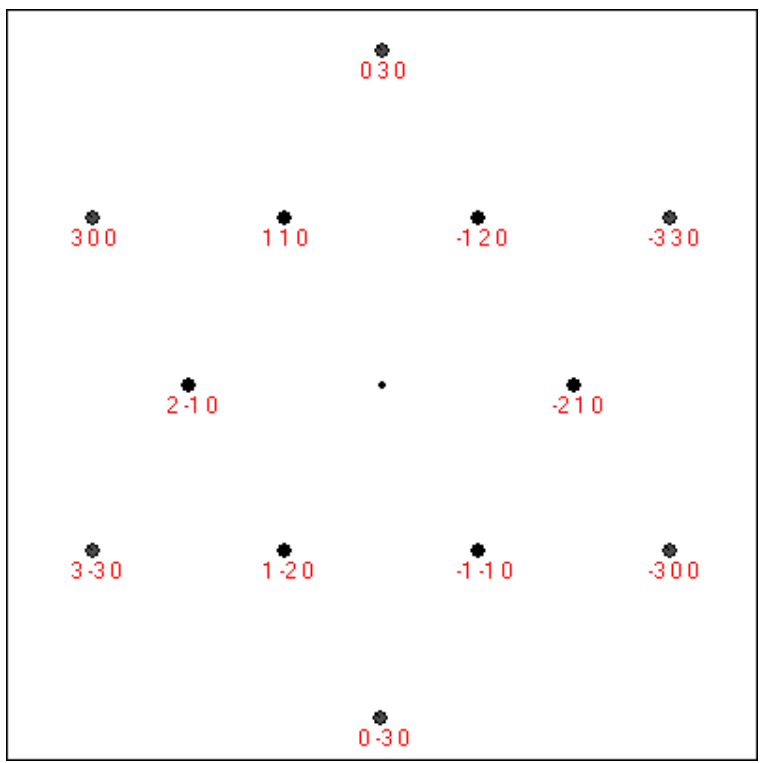

**Figure S5** Simulated SAED pattern of defect-free WN crystal.

**MoS$_2$ transistors with W$_5$N$_6$ and Au contacts**

To take advantage of the semimetallic properties of W$_5$N$_6$, we investigate the performance of W$_5$N$_6$ as a contact electrode to 2D semiconductors (*i.e.*, MoS$_2$). We fabricate MoS$_2$ transistors with W$_5$N$_6$ and Au contacts and compare their performance. Details of the fabrication of the heterostructures are presented in the Method section. We present the charge transport characteristics of W$_5$N$_6$-MoS$_2$ and Au-MoS$_2$ devices with application of different gate voltage (V$_{GS}$) in Figure S6. From Figure S6a, asymmetric I-V characteristic curves are observed under different V$_{GS}$, indicating the presence of Schottky barrier at the W$_5$N$_6$-MoS$_2$ and Au-MoS$_2$ interfaces. It is worth noting that with similar device dimensions, W$_5$N$_6$-MoS$_2$ device exhibit approximately ten times higher current compared to the Au-MoS$_2$ device. Moreover, n-type semiconductor behavior is observed from the transfer curve (Figure S6b) in both W$_5$N$_6$-MoS$_2$ and Au-MoS$_2$ devices. The W$_5$N$_6$-MoS$_2$ exhibits ~60 folds higher on-current compared to the Au-MoS$_2$ device. Both charge transport and transfer curve measurements indicate that a much lower contact resistance is present at the W$_5$N$_6$-MoS$_2$ interface compared to the Au-MoS$_2$ interface.

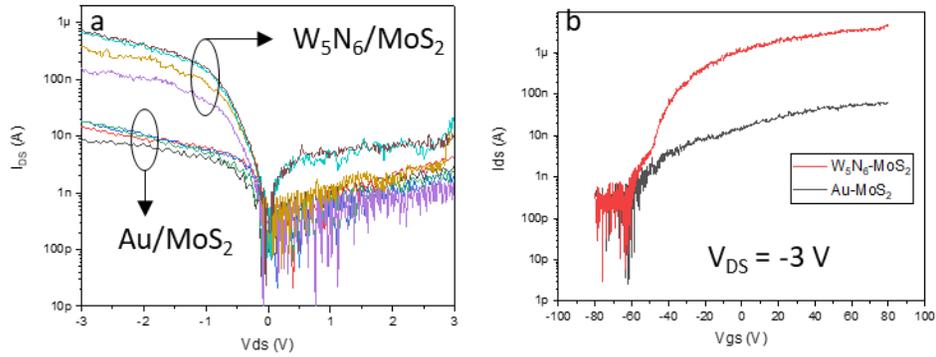

**Figure S6** Electrical measurements of $MoS_2$ FET device with $W_5N_6$ contact and Au contact. (a) Charge transport curves measured at different $V_{GS}$. The I-V curves are measured at $V_{GS}$= +20 V, +40 V, +60 V, +80 V, respectively from bottom to top using both $W_5N_6$ and Au contacts. (b) Transfer characteristic curves of the $MoS_2$ FET device with $W_5N_6$ and Au contacts.

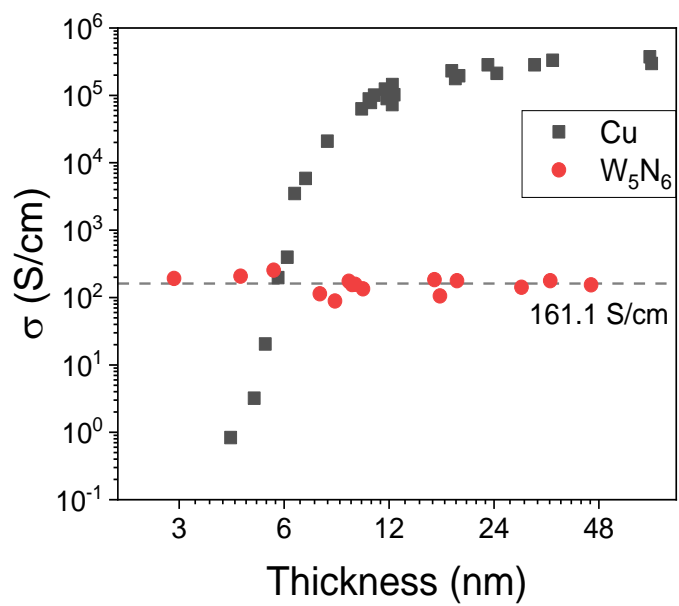

**Figure S7** Electrical conductivity of $W_5N_6$ (prepared by atomic substitution) and ultra-thin Cu films (prepared by vacuum deposition) at different thicknesses. Data of Cu is adapted from work reported by Schmiedl *et al.*[3]

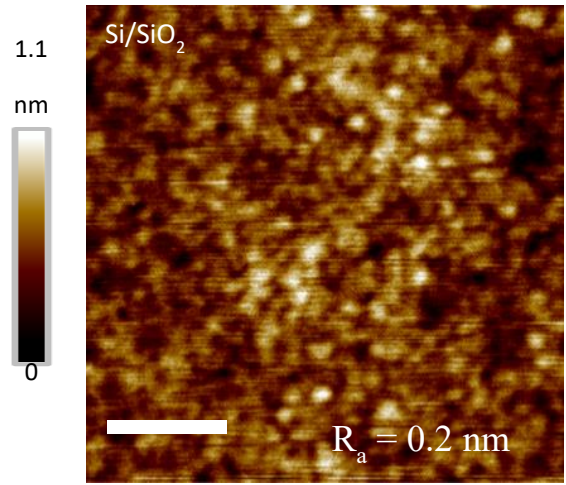

**Figure S8** AFM image of the typical SiO$_2$/Si substrate.

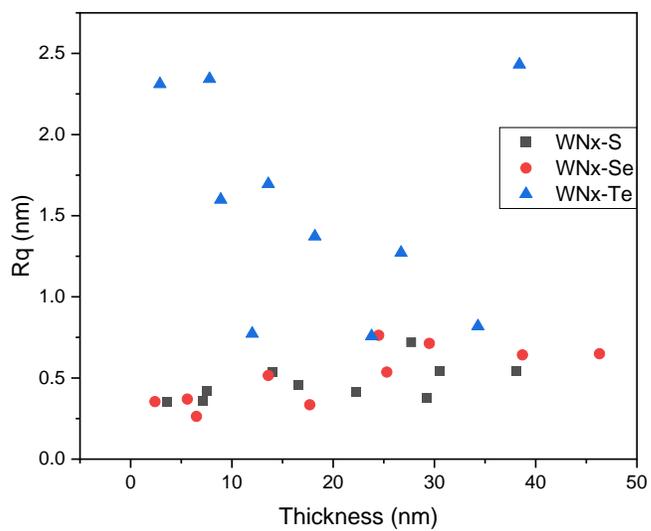

**Figure S9** Roughness of $WN_X$-S, $WN_X$-Se, and $WN_X$-Te at various thicknesses.

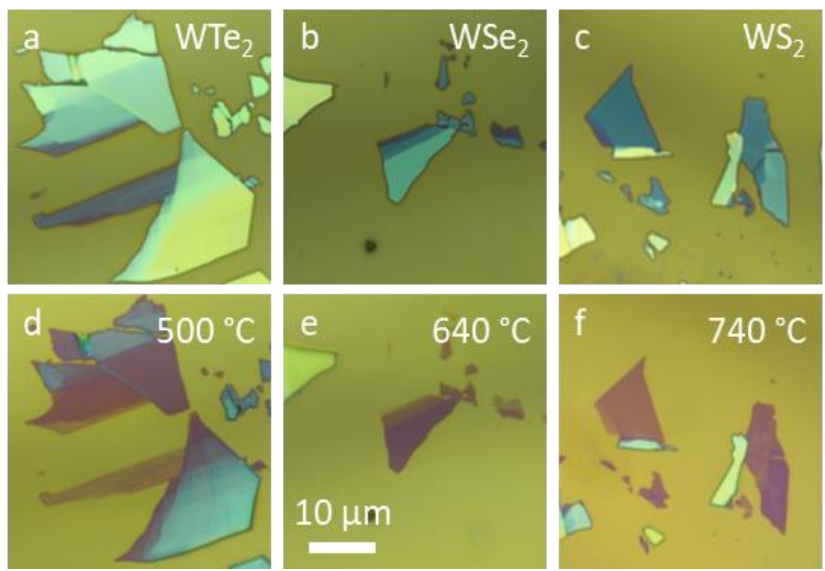

**Figure S10** Full conversion of each W-TMDs with the threshold temperature of conversion labeled on the images. All images share the same scale bar.

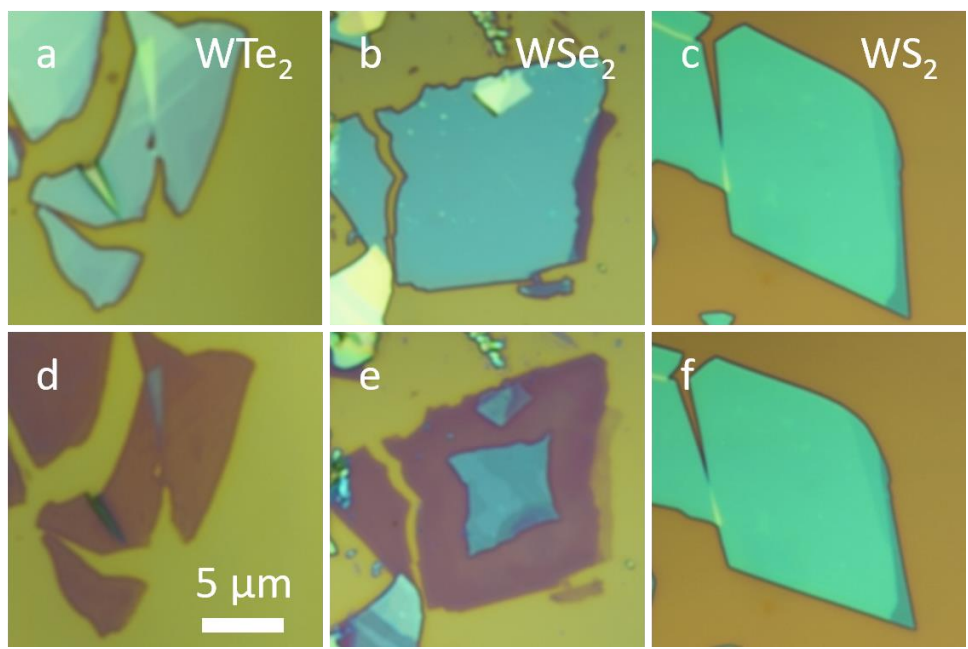

**Figure S11** W-TMDs reacted at 700 °C for 30 minutes in $NH_3$. (a-c) Optical images of precursor W-TMDs. (d-f) Optical images of flakes after reaction. $WTe_2$ is fully converted (d), $WSe_2$ is partially converted with obvious sign of the conversion initiating from the edge (e), and $WS_2$ shows no signs of conversion (f). All images share the same scale bar.

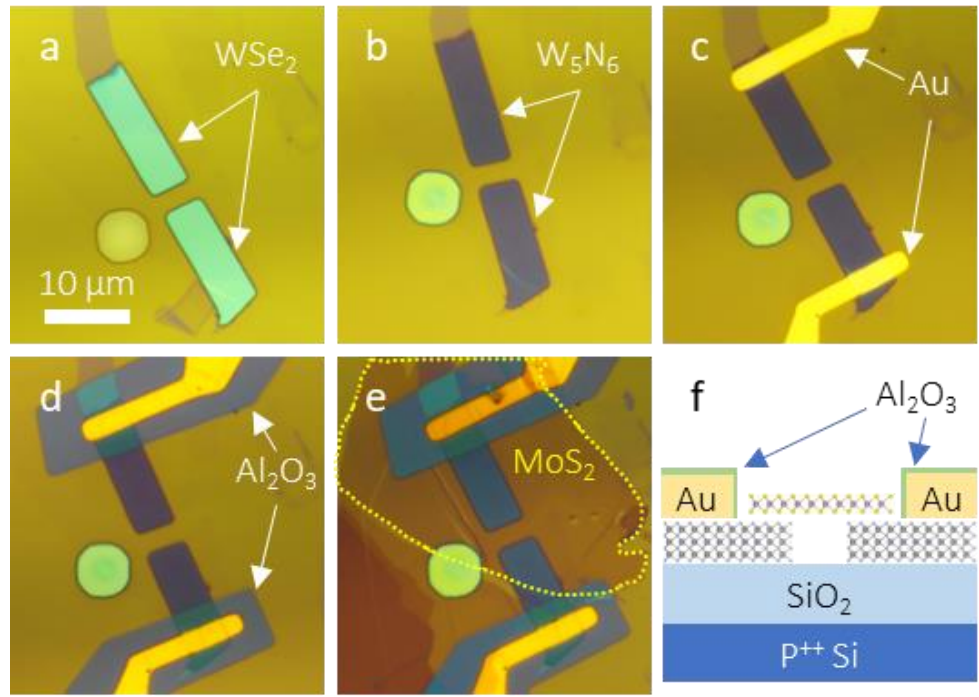

**Figure S12** Fabrication of $W_5N_6$-$MoS_2$ vdW heterostructure. (a) Etched $WSe_2$ stripes. (b) Etched $WSe_2$ converted into $W_5N_6$. (c) Deposit of Ti/Au electrode on $W_5N_6$. (d) Deposition of $Al_2O_3$ insulating layer to separate $MoS_2$ from Au. (e) Transfer and stack of monolayer $MoS_2$ onto the device to form $MoS_2$-$W_5N_6$ vdW contact. (f) Cross-sectional schematic of the device.

**Table S1** Lattice parameters extracted from SAED patterns taken at different locations of the $WSe_2$-$W_5N_6$ heterostructure. Each $d_{100}$ is measured from four diffraction patterns taken at different spots in the same region.

| Location | $d_{100}$ $WSe_2$ (Å) |
|---|---|
| $W_5N_6$ only | N/A |
| Interface | 2.95 ± 0.01 |
| $WSe_2$ only | 2.93 ± 0.01 |

**References**

(1) Zhao, Z.; Bao, K.; Duan, D.; Tian, F.; Huang, Y.; Yu, H.; Liu, Y.; Liu, B.; Cui, T. The Low Coordination Number of Nitrogen in Hard Tungsten Nitrides: A First-Principles Study. *Phys. Chem. Chem. Phys.* **2015**, *17* (20), 13397–13402.

(2) Xing, W.; Miao, X.; Meng, F.; Yu, R. Crystal Structure of and Displacive Phase Transition in Tungsten Nitride WN. *J. Alloys Compd.* **2017**, *722* (193), 517–524.

(3) Schmiedl, E.; Wissmann, P.; Finzel, H. U. The Electrical Resistivity of Ultra-Thin Copper Films. *Zeitschrift fur Naturforsch. - Sect. A* **2008**, *63* (10–11), 739–744.